\documentclass[12pt]{iopart}

\usepackage{latexsym}
\usepackage{iopams}  
\usepackage[dvipsnames]{xcolor}
\usepackage{comment}
\begin{document}

\title{An exploration of the black hole entropy in Gauss-Bonnet gravity via the Weyl tensor}

\author{Taha A. Malik \& Rafael Lopez-Mobilia}

\address{Department of Physics and Astronomy, University of Texas at San Antonio, San Antonio, Texas 78249, USA}
\ead{tahmalik@cantab.net, rafael.lopezmobilia@utsa.edu}
\vspace{10pt}
\begin{indented}
\item[]August 2020\end{indented}

\begin{abstract}
Penrose suggested that issues with the origin of the second law of thermodynamics and the remarkable homogeneous and isotropic nature of the universe on large scales can be resolved with a concept of an entropy for gravitational fields captured by the Weyl curvature tensor. This has led to various proposals for an entropy density for gravity constructed from the Weyl tensor. Most work investigating such entropy densities have been restricted to general relativity and little work has been done to them in modified theories of gravity. To remedy this, we investigate the simplest proposal for an entropy density in 5-dimensional Gauss-Bonnet gravity. 


\end{abstract}
%
%
%
%
%

\section{Introduction}
\label{002.intro}

\subsection{Entropy and the Generalized Second Law}
\label{002.EGSL}
The second law of thermodynamics states that in an isolated system the entropy, $S$, never decreases:
\begin{equation}\label{002.S>0}
\frac{dS}{dt}\geq 0.
\end{equation}
 Originally, entropy was considered as a new extensive thermodynamic property obeying Eq.(\ref{002.S>0}). But later, Ludwig Boltzmann showed that entropy could be understood as the logarithm of the number of different microstates, $W$, compatible with a given macrostate,
 \begin{equation}
S=\log(W).
\end{equation} \footnote{We have set the Boltzmann constant ($k_B=1$).} Boltzmann's approach informs our modern understanding of entropy and of the second law as fundamentally probabilistic. 

 In order for the second law of thermodynamics to be valid in the presence of black holes, Jacob Bekenstein proposed that black holes must possess nonzero entropy \cite{PhysRevD.7.2333}. Stephen Hawking later confirmed Bekenstein's idea and showed that black holes must have a temperature and radiate as black bodies \cite{Hawking1975}. The Bekenstein-Hawking entropy of black holes in general relativity is
\begin{equation}\label{002.EqW:BH}
S_{BH}=\frac{A}{4G\hbar}.
\end{equation}
The generalized entropy \cite{PhysRevD.9.3292} 
\begin{equation}
S_{gen}=S_{BH}+S_{out},
\end{equation}
where $S_{out}$ is the entropy outside the black hole and can be considered as the total entropy of a system when both black holes and matter are present. There is mounting evidence to suggest that classically and semi-classically $S_{gen}$ is never decreasing \cite{PhysRevD.85.104049,PhysRevD.82.124019}. This is known as the generalized second law (GSL) and supersedes the usual second law of thermodynamics. 

Many expect that fundamentally, the origin of $S_{gen}$ and GSL is statistical in nature, as with the usual notion of entropy. But perhaps its origin lies elsewhere and our notion of entropy needs modification. Any insights we can gain about the entropy of black holes and other spacetime geometries may provide hints towards the nature of quantum gravity. 

\subsection{Weyl curvature conjecture}
\label{002.WCC}

Understanding black hole thermodynamics and the realization that they have entropy has been important in understanding the status of the second law of thermodynamics and the need to generalize it (GSL). However, no universally agreed generalization for the entropy has been found for other spacetimes. Addressing this issue could help explain the apparent violation of the second law of thermodynamics on cosmological scales \cite{Page:1983aa,Penrose:aa1}. It seems that the universe today has evolved from a remarkably homogeneous and isotropic initial state to a non-equilibrium state which includes complex structures like stars and galaxies. Small perturbations in the initial state grew as a result of gravity and caused structures to form. For gravitational interactions to be made compatible with the second law of thermodynamics, Penrose suggested that gravitational fields should have an associated entropy \cite{Penrose:aa1}. Such a concept, if correct, could lead to a further generalization of the GSL and explain the homogeneous and isotropic nature of the universe.
 





The Weyl tensor $C_{\mu\nu\lambda \rho}$ is the trace-free part of the Riemann tensor and in an $n$-dimensional spacetime, the Riemann tensor can be decomposed into its trace-free and trace part:  
\begin{eqnarray}
R_{\mu\nu\lambda \rho}=\frac{1}{n-2}(g_{\mu\lambda}R_{\nu\rho} +g_{\nu\rho}R_{\mu\lambda} -g_{\nu\lambda}R_{\mu\rho} -g_{\mu\rho}R_{\nu\lambda}) \nonumber \\
+\frac{1}{(n-1)(n-2)}(g_{\mu\rho}g_{\nu\lambda} -g_{\mu\lambda}g_{\nu\rho})R+C_{\mu\nu\lambda \rho},
\end{eqnarray}
where $R_{\mu\nu\lambda \rho}$ is the Riemannn tensor, $R_{\mu\nu}$ is the Ricci tensor, $R$ is the Ricci scalar and $g_{\mu\nu}$ is the metric.

The Einstein field equations relate the curvature of spacetime to the matter content of spacetime, or more precisely
\begin{equation}\label{002.EE}
R_{\mu\nu}-\frac{1}{2}Rg_{\mu\nu}=8\pi GT_{\mu\nu},
\end{equation} 
where $G$ is the gravitational constant and we have set the speed of light $c=1$. The left hand side of this equation only involves the Ricci tensor and Ricci scalar and hence locally, the Weyl tensor is independent of the energy momentum tensor. It represents gravitational effects such as tidal effects and gravitational waves. Thus it seems plausible that information encoded in the Weyl tensor is related to the thermodynamics of gravity.

 The Weyl tensor also vanishes if and only if the spacetime is conformally flat in spacetimes with dimension $n>3$. Hence initially, in 4 or higher dimensions, when the universe was homogenous and isotropic, the Weyl tensor vanished and grew afterwards as complex structures formed. This motivated Penrose to conjecture (The Weyl curvature conjecture) that the low entropy content of the universe is related to the vanishing of the Weyl tensor \cite{Penrose:aa1, Goode_1985, 1985CQGra...2...99G, 1989NYASA.571..249P}.



 
To realize Penrose's conjecture, candidates for an entropy density for gravity have been proposed using the Weyl tensor \cite{Weyl2008,PhysRevD.62.044009,Barrow_2002,AGEP}. For example the simplest such entropy density below uses a scalar constructed from the Weyl tensor; 
\begin{equation}\label{002.P}
P^2=C^{\mu\nu\lambda\rho}C_{\mu\nu\lambda\rho}.
\end{equation}

Most work investigating such entropy densities have been restricted to general relativity and little work has been done to them in modified theories of gravity. To remedy this, we investigate Eq.(\ref{002.P}) using methods similar to those of ref\cite{Li2016}, but in 5-dimensional Gauss-Bonnet gravity.

\section{Gauss-Bonnet gravity}
\label{002.GBG}

Lovelock theories of gravity are a class of generalizations of general relativity in $n$-dimensional spacetime which give second order equations of motion for the metric \cite{doi:10.1063/1.1665613}. In $4$ dimensions, the Einstein field equations (Eq.(\ref{002.EE})) with a cosmological constant are the unique class of Lovelock theories. 

In $5$ dimensions, Gauss-Bonnet gravity is the most general Lovelock theory. Setting $c=1$, the action for Gauss-Bonnet gravity is given by \cite{boulware:1985aa,Clunan:2004aa}
\begin{equation}
\frac{1}{16\pi G} \int d^5x\sqrt{-g}(R+\alpha (R^2-4R^{\mu\nu}R_{\mu\nu}+R^{\mu\nu \lambda \rho}R_{\mu\nu \lambda \rho})),	
\end{equation}
where $\alpha$ is an arbitrary constant with dimensions of $[length]^2$. The $(R^2-4R^{\mu\nu}R_{\mu\nu}+R^{\mu\nu \lambda \rho}R_{\mu\nu \lambda \rho})$ term in $4$ dimensions is a topological term which can be removed from the action without affecting the equation of motion. But in $5$ dimensions, this term in general can not be removed. These higher-curvature terms may arise from quantum gravity corrections so we will  assume $\frac{\alpha}{l_p^2}\approx 1$.

We work with the following family of static vacuum (black hole) solutions for $5$-dimensional Gauss-Bonnet gravity \cite{boulware:1985aa,Clunan:2004aa}

\begin{equation}\label{002.Eq:g}
ds^2=-F(r)^2dt^2+\frac{1}{F(r)^2}dr^2+r^2d\Omega^2,	
\end{equation}
where $d\Omega^2$ is the metric on the $S^3$ sphere and 
	\begin{equation}\label{002.Eq:g2}
F^2(r)=k+\frac{r^2}{4\alpha}\left( 1-\sqrt{1+\frac{128\pi G\alpha M}{3 \Sigma_k r^{4}}+\frac{4\alpha\Lambda}{3}}  \right)
	\end{equation} with $\Sigma_k$ being the volume of the 3-sphere and $\Lambda$ the cosmological constant. $k$ can be either $-1$, $0$ or $1$ although $k=-1 $ and $k=0$ are only possible when $\Lambda<0$. $M$ parametrizes the family of solutions and is also the mass of the black hole. For simplicity, we now set $k=1$ and $\Lambda=0$, unless explicitly stated otherwise. Our calculations are independent of this choice. See \ref{002.AB} for more details.

To first order in $\alpha$, $F^2(r)=f^2(r)+\alpha\frac{M_1^2}{8r^6}$ with $f^2(r)=\left(1-\frac{M_1}{4r^2}\right)$ and the Weyl invariant to first order in $\alpha$ is 
	\begin{equation}\label{002.P2}
C_{\mu\nu\lambda \rho}C^{\mu\nu\lambda \rho}=\frac{9M_1^2}{2r^8}-\alpha \frac{21M_1^3}{r^{12}} +O(\alpha^2),
\end{equation} 
where $M_1=\frac{64\pi GM}{3\Sigma_k}$. 
	
	\section{Black hole entropy in Gauss-Bonnet Gravity}
\label{002.BHEGBG}
 
When the Einstein Hilbert action is modified, the expression for the black hole entropy also needs to be modified. For a stationary black hole's entropy in 5-dimensional Gauss-Bonnet gravity, this is given by \cite{Clunan:2004aa,Hawking1983} 
\begin{equation}\label{002.GB}
S_{BH}=\frac{R_5^3\Sigma_k}{4l_p^3}\left(1+\frac{12\alpha k}{R_5^2}\right)=\frac{A}{4l_p^3}\left(1+\frac{12\alpha k}{R_5^2}\right),
\end{equation}  
  where $A=R_5^3\Sigma_k$ is the surface area of the horizon at $r=R_5$ with $R_5$ being the largest root of $F^2(R_5)=0$ \footnote{We have explicitly reintroduced the parameter $k$ and $\Lambda$ for this expression. Also note that the black hole entropy is independant of $\Lambda$.}. The expression for black hole entropy changes because the behavior of the black hole horizon changes when the Einstein field equations are modified. This can be seen by considering the first law of black hole mechanics \cite{Clunan:2004aa} or via Wald's Noether charge method \cite{Wald:1993aa}.

In this paper, we will work in 5 dimensions and integrate the expression $\int C_{\mu\nu\lambda \rho}C^{\mu\nu\lambda \rho} dV_4$ on a spacelike hypersurface normal to the timelike Killing vector field to see if we recover the entropy of a black hole in Gauss-Bonnet gravity (Eq.(\ref{002.GB})). By dimensional analysis, the integral $\int C_{\mu\nu\lambda \rho}C^{\mu\nu\lambda \rho} dV_4$ to zeroth order in $\alpha$ can be written as 

\begin{equation}\label{002.n1}
	a\left(\frac{R_5}{l_p}\right)^{n_1}+\dots
\end{equation}

where $a$ and $n_1$ are constants to be determined. We only focus on the highest order term in $\frac{R_5}{l_p}$ with lower order terms contained in `\dots' which we regard as small for large black holes. Similarly, to first order in $\alpha$, the integral $\int C_{\mu\nu\lambda \rho}C^{\mu\nu\lambda \rho} dV_4$ can be written as

\begin{equation}\label{002.n2}
	b\frac{\alpha}{R_5^2}\left(\frac{R_5}{l_p}\right)^{n_2}+\dots
\end{equation}
where $b$ and $n_2$ are constants to be determined with lower order terms in $\frac{R_5}{l_p}$ ignored. For Eq.(\ref{002.P}) to pass as an entropy density in Gauss-Bonnet gravity, we must have $n_1=n_2=3$\footnote{Also, $b=12a$}. However, we show below that $n_2=7$.

\section{Dimensional Analysis}\label{002.DA} 

Fortunately, we can determine $n_1$ and $n_2$ directly through dimensional analysis. However, we provide a more detailed calculation in \ref{002.AA}. 

The integral $\int C_{\mu\nu\lambda \rho}C^{\mu\nu\lambda \rho} dV_4$ up to first order in $\alpha$ is

\begin{eqnarray}\label{002.ECI}
	 \int C_{\mu\nu\lambda \rho}C^{\mu\nu\lambda \rho} dV_4 &=2\pi^2\int_{l_p}^{R_5}\left(\frac{A}{r^8}-\frac{\alpha B}{r^{12}}\right)r^3g_{rr}dr \nonumber \\
	& =2\pi^2\int_{l_p}^{R_5}\left(\frac{A}{r^8}-\frac{\alpha B}{r^{12}}\right)\frac{r^3}{|F(r)|}dr. 
\end{eqnarray}
where $A=\frac{9M_1^2}{2}$ and $B=21M_1^3$ are constants independent of $r$. The classical theory of general relativity and generalizations like Gauss-Bonnet gravity are expected to break down at extremely small radii where we expect a quantum theory of gravity to have large effects. In particular, these theories are invalid at the spacetime singularity ($r=0$) so we bound the integral below by Planck's constant $l_p$ in Eq.(\ref{002.ECI}). In this case, we find that the highest order contribution in $\frac{R_5}{l_p}$ at zeroth and first order in $\alpha$ comes from the small radius (i.e $r \rightarrow l_p$)\footnote{In other words, $n_1$ and $n_2$ can be extracted from the integral near $r\approx l_p$} (see \ref{002.AA}). 

When $r \approx l_p$, we can approximate $F^2(r)$ from Eq.(\ref{002.Eq:g2}) as

\begin{equation}
F^2(r)=1-\frac{C}{r^2}+\frac{\alpha D}{r^6} + O(\alpha^2),
\end{equation} 
for some positive constants $C$ and $D$ which means that 
\begin{equation}
	\frac{r^3}{|F(r)|}\approx\frac{r^3}{|1-\frac{C}{r^2}+\frac{\alpha D}{r^6}|^{\frac{1}{2}}}\approx \frac{r^3}{(\frac{C}{r^2}-\frac{\alpha D}{r^6})^{\frac{1}{2}}}\approx C^{-\frac{1}{2}}\left(r^4+ \frac{\alpha D}{C}\right).
\end{equation}
Hence 
\begin{eqnarray}\label{002.Int}
	 \int C_{\mu\nu\lambda \rho}C^{\mu\nu\lambda \rho} dV_4 &\propto \int_{r\approx l_p}\left(\frac{A}{r^8}-\frac{\alpha B}{r^{12}}\right)\left(r^4+\frac{\alpha D}{C}\right)dr \nonumber \\ 
	 & \approx \int_{r\approx l_p} \left(\frac{A}{r^4}+\alpha\frac{(AD/C-B)}{r^8} \right)dr.
\end{eqnarray}
Thus the essential contribution to the integral coming from the $r\approx l_p$ piece gives
\begin{eqnarray}\label{002.EDBH}
	 \int C_{\mu\nu\lambda \rho}C^{\mu\nu\lambda \rho} dV_4 \propto J\left(\frac{R_5}{l_p}\right)^3+K \frac{\alpha}{R_5^2} \left(\frac{R_5}{l_p}\right)^7 
\end{eqnarray}
for some dimensionless constants $J$ and $K$. We plugged back the black hole radius $R_5$ in this expression to emphasize that the integral is dimensionless. This shows that $n_1=3$ and $n_2=7$. Since $n_2\neq3$, we find that the integral above does not reduce to the expected expression for black hole entropy in Gauss-Bonnet gravity (Eq.(\ref{002.GB})) at order $\alpha$. Thus Eq.(\ref{002.P}) is not a valid entropy density for Gauss-Bonnet gravity.

	\section{Discussion}
\label{002.DIS}

The second law of thermodynamics is one of the most important in physics because of its wide range of applicability. For it to be valid when a system includes a black hole, one has to assign an entropy to the black hole, $S_{BH}$. If the notion of black hole entropy can be extended to other spacetimes and cosmological settings, then the apparent violation of the second law of thermodynamics on cosmological scales may be resolved \cite{Page:1983aa,Penrose:aa1}. Penrose conjectured that the initially  low  entropy of the universe is related to the vanishing of the Weyl tensor. Motivated by this, various proposals for an entropy density constructed from the Weyl tensor have been made to realize this conjecture \cite{Weyl2008,PhysRevD.62.044009,Barrow_2002,AGEP}. In $5$ dimensions, the simplest scalar constructed from the Weyl tensor was proposed as an entropy density (see Eq.(\ref{002.P})) and investigated in ref\cite{Li2016}. They found that this entropy density recovers the Bekenstein-Hawking entropy of black holes. 


When we replicated this investigation in Gauss-Bonnet gravity, a generalization of general relativity, the proposed entropy density failed to recover the standard black hole entropy at order $O(\alpha)$. By dimensional analysis, we see that this is due to the term $(-21\alpha\frac{M1^3}{r^{12}})$ in Eq.(\ref{002.P2}) which ultimately leads to $n_2=7\neq3$ in Eq.(\ref{002.n2}). For the entropy density to reproduce the black hole entropy, we require $n_1=n_2=3$. In other words, we require the entropy density $s$ to take the following form;

\begin{equation}\label{002.sd}
	s=\frac{L}{r^8}+\frac{N\alpha}{R_5^2r^8}+O(\alpha^2).
\end{equation}
where $L$ and $N$ are some constants. Other candidates for an entropy density multiply Eq.(\ref{002.P}) by scalars constructed from other curvature terms \cite{Weyl2008, PhysRevD.62.044009}. An example of such a candidate is 
\begin{equation}
	P'^2=C^{\mu\nu\lambda\rho}C_{\mu\nu\lambda\rho}/R_{\delta}^{\gamma}R_{\gamma}^{\delta}.
\end{equation}
By dimensional analysis, none of these candidates can change the form of the entropy density from Eq.(\ref{002.P2}) to Eq.(\ref{002.sd}). 

In Gauss-Bonnet gravity, the Einstein field equations are modified so that the Weyl tensor couples to the energy momentum tensor and is not necessarily locally independent of the matter content in spacetime. This and our investigation above suggest that the Weyl tensor may need to be replaced in alternative theories of gravity with some other tensor constructed from the Riemann tensor which is locally independent of the matter content. 


In vacuum with the cosmological constant set to zero ($\Lambda=0$), only the gravitational degrees of freedom are involved, so we expect the Riemann tensor to represent only these degrees of freedom. Additionally, in general relativity, vacuum solutions with $\Lambda=0$ lead to $R_{\mu\nu}=0$ and so $C_{\mu\nu\lambda\rho}=R_{\mu\nu\lambda\rho}$. Thus, in vacuum with $\Lambda=0$, 
\begin{equation}
	P^2=C^{\mu\nu\lambda\rho}C_{\mu\nu\lambda\rho}=R^{\mu\nu\lambda\rho}R_{\mu\nu\lambda\rho}.
\end{equation}

In Gauss-Bonnet gravity, even if the Weyl tensor needs to be replaced by curvature terms locally independent of the matter content, we still expect the Riemann tensor to represent only the gravitational degrees of freedom in a vacuum with $\Lambda=0$. Thus one may expect that in vacuum with $\Lambda=0$, 
\begin{equation}
P^2=\tilde{C}^{\mu\nu\lambda\rho}\tilde{C}_{\mu\nu\lambda\rho}=R^{\mu\nu\lambda\rho}R_{\mu\nu\lambda\rho}
\end{equation}
where $\tilde{C}^{\mu\nu\lambda\rho}$ is the conjectured curvature term locally independent of the matter content replacing the Weyl tensor. However, for the black hole solutions in Gauss-Bonnet gravity with $\Lambda=0$, we find that $\int R_{\mu\nu\lambda \rho}R^{\mu\nu\lambda \rho} dV_4=\int C_{\mu\nu\lambda \rho}C^{\mu\nu\lambda \rho} dV_4$ up to first order in $\alpha$ so our results don't change. Thus, the issues resulting from dimensional analysis would still need to be tackled head on.

Another issue with Eq.(\ref{002.P2}) is that it leads to an expression for the black hole entropy, which is independent of the $k$ parameter (see \ref{002.AB}). This is incompatible with the standard black hole entropy, Eq.(\ref{002.GB}), as it depends on $k$. 

We expect that some of these issues could be resolved if the entropy density is a sums of scalars built from curvature terms locally independent of the matter content. For example, using a Bel-Robinson-like tensor for Gauss-Bonnet to find a ``super-energy-momentum tensor" \cite{AGEP} may help in building the curvature terms that would replace its Weyl tensor counterpart. 

Additionally any curvature term locally independent of the matter content that is to replace the Weyl tensor in Gauss-Bonnet gravity would need to be checked that it satisfies similar properties to those discussed in section (\ref{002.WCC}), which motivated the Penrose Weyl curvature conjecture in the first place. We leave this to future work.

Finally, we note that the expression for the entropy density in Eq.(\ref{002.P2}) includes a negative contribution proportional to $\alpha$ which dominates for $l_p<r\lessapprox(\alpha^{\frac{1}{2}}R_5)^{\frac{1}{2}}\approx (l_pR_5)^{\frac{1}{2}}$ where the upper bound can be large for large black holes. This further shows that the entropy density needs modification in Gauss-Bonnet gravity since a negative entropy density is not physically meaningful. We hope that future work on finding a suitable replacement to Eq.(\ref{002.P2}) will resolve this issue.

One may worry that this analysis was artificially restricted to 5-dimensions. However, using similar methods, one could test other proposals in other dimensions, provided the constructed entropy density via the Weyl tensor has the correct physical dimension. For instant, one could investigate $\tilde{P}^{(N)}:=(C_{abcd}C^{abcd})^{\frac{N-1}{4}}=P^{\frac{N-1}{2}}$ instead of Eq.(\ref{002.P}) as an entropy density in $N$-dimensions. To test Gauss-Bonnet gravity, a 5 or higher-dimensional spacetime is required for a non-trivial test because in lower dimensions Gauss-Bonnet gravity reduces to general relativity. In higher dimensions, non-trivial tests could also be conducted in other Lovelock theories of gravity, which are further generalizations of general relativity and Gauss-Bonnet gravity. We expect similar issues to arise in these cases as well.

\section*{ACKNOWLEDGMENTS}

We would like to thank the anonymous referees for constructive feedback and helpful comments. In particular, the dimensional analysis argument used in section \ref{002.DA} is based on suggestions made by one of the referees.

\appendix
\section{\\Evaluation of $\int C_{\mu\nu\lambda \rho}C^{\mu\nu\lambda \rho} dV_4$}\label{002.AA}

Evaluating $\int C_{\mu\nu\lambda \rho}C^{\mu\nu\lambda \rho} dV_4$ using the expression in Eq.(\ref{002.P2}), we find to first order in $\alpha$ that
 \begin{eqnarray}
 &\int C_{\mu\nu\lambda \rho}C^{\mu\nu\lambda \rho} dV_4 = \int C_{\mu\nu\lambda \rho}C^{\mu\nu\lambda \rho}r^3|g_{rr}|^{\frac{1}{2}} drd\Omega_3 \nonumber \\
 &= 2\pi^2\int_{l_p}^{R_5} C_{\mu\nu\lambda \rho}C^{\mu\nu\lambda \rho}r^3|g_{rr}|^{\frac{1}{2}} dr \nonumber \\
  &=2\pi^2\int_{l_p}^{R_5} C_{\mu\nu\lambda \rho}C^{\mu\nu\lambda \rho}\frac{r^3}{|F(r)|}dr  \nonumber \\
  &\approx 2\pi^2\int_{l_p}^{R_5} C_{\mu\nu\lambda \rho}C^{\mu\nu\lambda \rho}r^3\left|\frac{1}{(1-\frac{M_1}{4r^2})^\frac{1}{2}}-\alpha \frac{M_1^2}{4r^4(-M_1+4r^2)(1-\frac{M_1}{4r^2})^\frac{1}{2}}\right|dr \nonumber  \\
  \nonumber  \\
 & \approx \Bigg [ 2\pi^2\int_{l_p}^{R_5} \frac{9M_1^2}{2r^8}r^3\frac{dr}{(-1+\frac{M_1}{4r^2})^\frac{1}{2}}+2\pi^2\int_{R_5}^{\infty} \frac{9M_1^2}{2r^8}r^3\frac{dr}{(1-\frac{M_1}{4r^2})^\frac{1}{2}}\nonumber \\ \nonumber \\
 &-2\alpha\pi^2\int_{l_p}^{R_{5}} \frac{dr}{(-1+\frac{M_1}{4r^2})^\frac{1}{2}}\left(\frac{9M_1^2}{2r^8}r^3 \frac{M_1^2}{4r^4(-M_1+4r^2)} +\frac{21M_1^3}{r^{12}}r^3 \right) \nonumber \\ \nonumber \\ 
&-2\alpha\pi^2\int_{R_{5}}^{\infty} \frac{dr}{(1-\frac{M_1}{4r^2})^\frac{1}{2}} \left(\frac{9M_1^2}{2r^8}r^3 \frac{M_1^2}{4r^4(-M_1+4r^2)} +\frac{21M_1^3}{r^{12}}r^3\right)\Bigg ] \nonumber \\ & + O(\alpha^2) 
\end{eqnarray}
where $\int d\Omega_3=2\pi^2$ is the solid angle in 3-dimensional space and we made the following approximation
\begin{eqnarray}\label{002.W2}
&|\frac{1}{(1-\frac{M_1}{4r^2})^\frac{1}{2}}-\alpha \frac{M_1^2}{4r^4(-M_1+4r^2)(1-\frac{M_1}{4r^2})^\frac{1}{2}}| \nonumber \\ 
&\approx \frac{1}{|1-\frac{M_1}{4r^2}|}^\frac{1}{2}\left(1-\alpha \frac{M_1^2}{4r^4(-M_1+4r^2)} \right).
\end{eqnarray}
This approximation breaks down near $r=R_5$. We find that our conclusions are unaffected by this breakdown although we briefly discuss below how to treat the integral near this point.




At zeroth order in $\alpha$, $\int C_{\mu\nu\lambda \rho}C^{\mu\nu\lambda \rho} dV_4$ approximately reduces to the same expression as given in Eq.(\ref{002.W}), as expected. The integral at order $O(\alpha)$ does not converge since the integral $\int \left(1-\frac{M_1}{4r^2}^2\right)^{-\frac{3}{2}}$ is divergent near $R_5$ (the black hole horizon). But this point is where the approximation we used in Eq.(\ref{002.W2}) breaks down. One could evaluate the integral above near $R_5$ by considering this piece separately and replacing Eq.(\ref{002.W2}) with a better approximation. However, since we only need to pick out the piece to highest order in $\frac{R_5}{l_p}$, we do not evaluate the integral for all these pieces and instead focus on the piece that gives the $n_2$ contribution:
\begin{eqnarray}\label{002.W3}
& 2\alpha \pi^2 \int_{1+\epsilon}^{\frac{R_5}{l_p}}\left(\frac{9M_1^4o^{9}}{32R_5^{10}(-1+\frac{M_1o^2}{4R_5^2})^\frac{3}{2}}- \frac{21M_1^3o^7}{R_5^5(-1+\frac{M_1o^2}{4R_5^2})^\frac{1}{2}}do\right) \rightarrow \nonumber  \\
& \frac{2\alpha}{R_5^2} \pi^2 \int^{\frac{R_5}{l_p}}\left(\frac{9M_1^4o^{9}}{32R_5^{8}(\frac{M_1o^2}{4R_5^2})^\frac{3}{2}}-\frac{21M_1^3o^7}{R_5^3(\frac{M_1o^2}{4R_5^2})^\frac{1}{2}}do \right) \nonumber \\ 
& \approx K\frac{\alpha}{R_5^2}\left(\frac{R_5}{l_p}\right)^7, 
\end{eqnarray}
\footnote{A small term $\epsilon$ was introduced to avoid the coordinate singularity.} where we substituted $r$ with $r=\frac{R_5}{o}$ and $K$ is some constant. 

Note that the essential contribution from the integral comes from $o \propto 1/l_p$ or for $r$ small.

\section{\\ General $k$ and $\Lambda$}\label{002.AB}

For general $k$ and $\Lambda$, we find that 
\begin{eqnarray}
	C_{\mu\nu\lambda \rho}C^{\mu\nu\lambda \rho}&=\frac{(-2kr^2+2r^2+3M_1)^2}{2r^8} \nonumber \\
	& -\alpha\frac{6(\Lambda r^4+\frac{7M_1}{2})(M_1-\frac{2(k-1)r^2}{3})M_1}{r^{12}}
	\end{eqnarray}
and
\begin{eqnarray}
	\frac{1}{F(r)}&=\frac{2}{\sqrt{4 k-\frac{1}{3} 2 \Lambda r^2-\frac{M}{r^2}}} \nonumber \\
&+\alpha\frac{ \left(2 \Lambda r^4+3 M\right)^2}{2 r^4 \sqrt{36 k-6 \Lambda r^2-\frac{9 M}{r^2}} \left(-12 k r^2+2 \Lambda r^4+3 M\right)}
	\end{eqnarray}	
	to first order in $\alpha$. However the essential contribution to the integrals in section \ref{002.DA} comes from small radius (i.e $r\rightarrow 0$). In this case, we find that for small $r$, the expression above approximately reduce to the case when $k=1$ and $\Lambda$ (i.e. independent of $k$ and $\Lambda$).

\section*{References}
\bibliographystyle{unsrt}
\bibliography{references}

\end{document}